\documentclass{article}
\usepackage[utf8]{inputenc}

\usepackage{amsmath}
\usepackage{amssymb}

\usepackage{cite}
\usepackage{float}  
\usepackage{hyperref} 

\usepackage{lineno}

\usepackage{amsthm}
 
\theoremstyle{definition}

\usepackage{setspace}

\usepackage{epsfig}
\usepackage{amsthm}
\usepackage{lscape,graphicx}
\usepackage{array}
\usepackage{caption}
\usepackage{setspace}
\usepackage{algorithm}

\usepackage{fullpage}
\usepackage{centernot}
\usepackage{bbm}
\usepackage{chemarr}

\begin{document}

\title{\LARGE \bf
Discrete Flux and Velocity Fields of Probability and Their Global Maps in Reaction Systems}

\date{}

\maketitle

\begin{center}
      {Anna Terebus$^{1}$, Chun Liu$^{2}$,  and Jie Liang$^{1,*}$},\\[.1cm] 
    
    \end{center}

\noindent
\thanks{$^{1}$ Department of Bioengineering, University of Illinois at Chicago, Chicago IL, 60607, USA\\
\thanks{$^{2}$ Department of Applied Mathematics, Illinois Institute of Technology, Chicago IL, 60616, USA}\\
\thanks{$^{*}$ Corresponding author, \tt{jliang@uic.edu}}\\

\maketitle
\thispagestyle{empty}

\begin{abstract}
 Stochasticity plays important roles in reaction systems. Vector fields
of probability flux and velocity characterize time-varying and
steady-state properties of these systems, including
high probability paths, barriers, checkpoints among different stable
regions, as well as mechanisms of dynamic switching among them.
However, conventional fluxes on continuous space are ill-defined and
are problematic when at boundaries of the state space or when copy
numbers are small.
By re-defining the derivative and divergence operators based on the
 discrete nature of reactions, we introduce new formulations of
 discrete fluxes.  Our flux model fully accounts for the discreetness
 of both the state space and the jump processes of reactions.
The reactional discrete flux satisfies the continuity equation and
 describes the behavior of the system evolving along directions of 
 reactions.
The species discrete flux directly describes the dynamic behavior in the state
space of the reactants such as the transfer of probability mass.
With the relationship between these two fluxes specified, we show how to
construct time-evolving and steady-state global flow-maps of
probability flux and velocity in the directions of every species at
every microstate, and how they are related to the outflow and inflow
of probability fluxes when tracing out reaction trajectories.
We also describe how to impose proper conditions enabling exact quantification of flux and velocity in the boundary regions, without
 the difficulty of enforcing artificial reflecting conditions. We illustrate the computation of probability flux and velocity
using three model systems, namely, the birth-death process, the
bistable Schl\"ogl model, and the oscillating Schnakenberg model.
\end{abstract}

\maketitle

\section{INTRODUCTION}
\label{Introduction} 
Biochemical reactions in cells are
intrinsically stochastic~\cite{mcadams1999sa, mcadams1997stochastic,
  kaern2005stochasticity, shahrezaei2008colored}.  When the
concentrations of participating molecules are small or the differences
in reaction rates are large, stochastic effects become
prominent~\cite{elowitz2002stochastic, swain2002intrinsic,
  kaern2005stochasticity, vellela2007quasistationary}.  Many stochastic
models have been developed to gain understanding of these reaction
systems~\cite{gillespie1977exact, vo2017adaptive, wolf2010solving,burrage2006krylov, cao2016state}.  These models either generate time-evolving landscapes of
probabilities over different microstates~\cite{vo2017adaptive, wolf2010solving,burrage2006krylov, cao2016state}, 
or generate trajectories along
which the systems travel~\cite{gillespie1977exact, cao2013adaptively}. Vector fields of probability flux and probability velocity 
are also of significant interest, as they can further
characterize time-varying properties of the reaction systems, including that of the non-equilibrium steady
states~\cite{zia2007probability, li2011landscape, zhang2012stochastic,
  wang2008potential, xu2012energy, yuan2013exploring}.  
For example, determining the
probability flux can help to infer the mechanism of dynamic switching among different
attractors~\cite{strasser2012stability, margaret2015dna}. Quantifying the probability flux can also help to characterize 
the departure of non-equilibrium reaction systems from detailed balance~\cite{zhang2012stochastic, de2014role, bianca2015evaluation}, and can 
help to identify barriers and checkpoints between different
stable cellular states~\cite{li2014landscape}. Computing probability fluxes and
velocity fields has found applications in studies of stem cell differentiation~\cite{wang2010potential}, cell cycle~\cite{li2014landscape}, and cancer
development~\cite{li2014quantifying, tang2017potential}.

 Models of probability fluxes and velocities in well-mixed mesoscopic chemical reaction systems have been the
  focus of many studies~\cite{xu2012energy,
  wang2008potential, li2014landscape, bianca2015evaluation, de2014role,  sjoberg2009fokker, mou1986stochastic,  schmiedl2007stochastic, xu2014energetic,schultz2008extinction,
  strasser2012stability}. They are often based on the formulation of the Fokker-Planck and the Langevin equations, both involving the
assumption of Gaussian noise of two moments~\cite{xu2012energy,
  wang2008potential, yuan2013exploring, tang2014nonequilibrium, li2014landscape, bianca2015evaluation}.  However,
these models are not valid when copy numbers of molecular species are
small~\cite{ van2007stochastic, sjoberg2009fokker, grima2011accurate,
  duncan2015noise}, as they do not provide a full account of the
stochasticity of the system~\cite{gillespie2000chemical,
  baras1996microscopic, van2007stochastic, sjoberg2009fokker,
  grima2011accurate, duncan2015noise}.  For example, the Fokker-Planck
model fails to capture multistability in gene regulation networks with
slow switching between the ON and the OFF
states~\cite{duncan2015noise}. These models are also of inadequate
accuracy when systems are far from
equilibrium~\cite{grima2011accurate}.  Moreover, solving the systems
of partial differential equations resulting from the Fokker-Planck and
Langevin Equations requires explicit boundary conditions for
states where one or more molecular species have zero
copies~\cite{xu2012energy}. 
These boundary conditions are ill-defined in the context of Gaussian
noise~\cite{zhou2016construction} and are difficult to impose using the Fokker-Planck/Langevin
formulation, or any other continuous models, as reactions cannot occur on boundary states when one or more reactants
are exhausted.

Several discrete models of probability flux and velocity based on continuous-time Markov jump processes associated
with the firing of reactions have also been
introduced~\cite{mou1986stochastic,
  schmiedl2007stochastic, schultz2008extinction,
  strasser2012stability}.  However, these models have limitations. 
   The models developed
in~\cite{schultz2008extinction, strasser2012stability} account only
for outflow fluxes. While the probability of transition to a
subsequent microstate after a reaction jump is accounted for, the
inflow flux describing the probability of transition into the current
microstate from a previous state is not explicitly considered. The work in~\cite{bazzani2012bistability} studies the phosporylation and dephosophorylation process. It introduces a formulation of discrete flux based on a forward finite difference operator. However, this is 
only applicable to
this special system of simple single-species reactions, where there is no mass exchange between the two different molecular types. The
models developed in~\cite{mou1986stochastic, schmiedl2007stochastic} are limited to analysis of single reactional
trajectories. In addition,  the probability flux is often assumed to be associated with
reactions that are reversible~\cite{horowitz2014thermodynamics}.
While these models offer an
in-the-moment view on how probability mass moves in the system by following trajectories generated from reaction events, 
they do not offer a global picture of the time-evolving probability flux at a specific time or at fixed locations in the state space.  
To construct the global flow-map of discrete probability flux and velocity,
proper formulations of discrete
flux and velocity, as well as methods to quantify the discrete forward and
backward flux between every two states connected by reactions are required.
  
In this study, we introduce
the appropriate formulations of discrete flux and discrete velocity for arbitrary mesoscopic reaction systems. We redefine  the derivative operator and discrete divergence based on the 
discrete nature of chemical reactions. The discreetness of both the state space and the jump processes of reactions 
is taken into consideration, with the
discrete version of the continuity equation satisfied. 
Our approach allows the quantification
of probability flux and velocity at every microstate,  as well as the ability in tracing out the
outflow probability fluxes and the inflow fluxes as reactions proceeds. 
In addition,
 proper boundary conditions are
imposed so vector fields of flux and velocity can be exactly computed anywhere in the discrete state space, without the difficulty of enforcing artificial reflecting conditions at the
boundaries~\cite{ceccato2018remarks}.
Our method can be used to 
exactly quantify transfer of probability mass and to
construct the global flow-map of the probability flux in all allowed directions of reactions 
over the entire state space. 
Results computed using our model can provide useful characterization of the
dynamic behavior of the reaction system, including the high probability paths
along which the probability mass of the system evolves, as well as
properties of their non-equilibrium steady states.

The accurate construction of the discrete probability flux, velocity,
and their global flow-maps requires the accurate calculation of the time-evolving probability landscape of
the reaction networks.
Here we employ the
recently developed ACME method~\cite{cao2016state, cao2016accurate}
to compute the exact time-evolving probability landscapes of networks by solving the underlying
discrete Chemical Master Equation (dCME).  This eliminates potential problems arising from inadequate
sampling, where rare events of low probability are
difficult to quantify using techniques such as the stochastic simulations algorithm (SSA)~\cite{gillespie1977exact, daigle2011automated, cao2013adaptively}.  

This paper is organized as follows. We first briefly discuss
the theoretical framework of reaction networks and discrete Chemical Master Equation. We then introduce
the concept of ordering of the microstates of the system, the definitions of discrete derivatives and
divergence, as well as flux and velocity on a discrete state space.
We further illustrate how time-evolving probability flux and velocity fields can be computed for three classical
systems, namely, the birth-death process~\cite{allen2010introduction,cao2016state}, the bistable Schl\"ogl model~\cite{schlogl1972chemical,cao2013adaptively}, and the oscillating Schnakenberg system~\cite{qian2002concentration,  cao2010nonlinear, xu2012energy}.

\section{Models and Methods}
\label{Models and Methods}

\subsection{Microstates, Probability, Reaction and Probability Vector}
\label{Microstates, Probability, Reaction and Probability Vector}

\textbf{Microstate and state space.}
We consider a well-mixed biochemical system with constant volume and
temperature. It has $n$ molecular species $X_i$,
$i = {1, \ldots ,n} $, which participate in $m$
reactions $R_k$, $k =  {1, \ldots ,m} $.
The \textit{microstate $\mathbf x(t)$} of the system at time $t$  is a column
vector of copy numbers of the molecular species: $\mathbf x(t) \equiv (x_1(t), x_2(t), \ldots , x_ n(t) )^T \in \mathbb{Z}_{+}^n$, where all values 
are non-negative integers. All the microstates that the system can reach 
form \textit{the state space} $\Omega = \{ \mathbf x(t) | t \in (0,\infty ) \}$.
The size of the state space is denoted as $\left| \Omega  \right|$.

\textbf{Probability and probability landscapes.} The probability of the system to be at a 
 particular microstate $\mathbf x$ at time $t$ is denoted as $p(\mathbf x, t) \in \mathbb{R}_{[0,1]}$. The probability
surface or landscape $\mathbf p (t)$ over the state space $\Omega$ is denoted as 
$\mathbf {p}(t) = \{p(\mathbf x, t) | \mathbf x \in  \Omega ) \}$.

\textbf{Reaction, discrete increment, and reaction direction.}
 \textit{A reaction} $R_k$ takes the general form of 
$$R_k: c_{1_{k}}X_1 +\cdots+ c_{n_{k}}X_n 
\stackrel{r_k}{\rightarrow} c'_{1_{k}}X_1 +\cdots+ c'_{n_{k}}X_n,$$
so that $R_k$ brings the system from a microstate $\mathbf x$ to $\mathbf x+\mathbf s_k$, where the stoichiometry vector
\begin{equation*}
\mathbf s_{k} \equiv (s_{k}^{1}, \ldots ,s_{k}^{n}) \equiv (c'_{1_{k}}-c_{1_{k}},\,\ldots \,,c'_{n_{k}}-c_{n_{k}})
\end{equation*}
gives the unit vector of \textit{the discrete increment} of reaction $R_k$. 
$\mathbf s_{k}$ 
also defines \textit{the direction of the reaction} $R_k$.
In a well-mixed mesoscopic system, the reaction propensity function $A_k(\mathbf x)$ is 
determined by the product of the intrinsic reaction rate $r_k$ and the combinations of 
relevant reactants in the current microstate $\mathbf x$: 
$$A_k (\mathbf x ) = r_k \prod\limits_{l = 1}^n {\left( {\begin{array}{*{20}c}
   {x_l }  \\
   {c_{lk} }  \\
\end{array}} \right)}.$$

\textbf{Discrete Chemical Master Equation and boundary states.} The discrete Chemical Master Equation (dCME) 
is a set of linear ordinary differential equations 
describing the changes of probability over 
time at each miscrostate of the system~\cite{mcquarrie1967stochastic, gillespie1977exact, cao2008optimal, cao2010probability}.  The dCME for an arbitrary microstate $\mathbf x=\mathbf x(t)$ can be written in the general form as: 
\begin{equation}
\frac{\partial p(\mathbf x, t)}{\partial t} =\sum\limits_{k=1}^m {  \normalsize [
A_{k} (\mathbf x-\mathbf s_k)  p (\mathbf x-\mathbf s_k,t)}-A_{k} (\mathbf x)  p (\mathbf x, t) \normalsize ], \quad  \mathbf x-\mathbf s_k, \, \mathbf x \in \Omega.
\label{eqn:dcme1}
\end{equation}

It is possible that only a subset or none of the permissible reactions can occur at a particular state $\mathbf x$ if it is at the boundary of the state space
$\Omega$, where the number of reactants is inadequate.
Specifically, we define \textit{the boundary states} $\partial \Omega_k$ for reaction $k$  as the states where reaction $R_k$ cannot happen:
\begin{eqnarray}
\partial \Omega_k  \equiv  \normalsize \{ {\mathbf x =  
(x_1,\ldots,x_i,\ldots,x_n)|\quad \textrm{there exist} \quad i:} x_i< c_{i_{k}} \normalsize \}.
\label{eqn:bc}
\end{eqnarray}

We define the overall boundary states as 
$
\partial \Omega \equiv \bigcup\limits_{k=1}^{m}  \partial \Omega_k. 
$

\textbf{Reactional probability vector and its time-derivative.}
We can consider each of the $k$-th reactions separately and decompose the right hand side of Eq.\,(\ref{eqn:dcme1}) into $m$ components, one for each reaction, $k=1 \ldots m$:
\begin{equation}
\frac{\partial p_k(\mathbf x, t)}{\partial t} = A_{k} (\mathbf x-\mathbf s_k)  p (\mathbf x-\mathbf s_k,t) - A_{k} (\mathbf x)  p (\mathbf x, t).
\label{eqn:dcme2}
\end{equation}
${{\partial p(\mathbf x,t)} \mathord{\left/ {\vphantom {{\partial  p(x,t)} {\partial t}}} \right. \kern-\nulldelimiterspace} {\partial t}}$ in Eq.(\ref{eqn:dcme1}) therefore can also be written as:
\begin{equation*}
\frac{\partial p(\mathbf x, t)}{\partial t} = \sum\limits_{k=1}^m \frac {\partial p_k(\mathbf x, t)}{\partial t}.
\end{equation*}
Any of the $m$ reactions can alter the value of $p(\mathbf x, t)$ 
as specified by Eq.(\ref{eqn:dcme2}). While the probability $p(\mathbf x, t)$ is a scalar,
we define \textit{the reactional probability vector} $\mathbf p(\mathbf x, t)$
such that 
\begin{equation}
\mathbf p(\mathbf x, t)=
(p_1(\mathbf x, t), \ldots , p_m(\mathbf x, t)) \in \mathbb{R}^m,
\label{eqn:vprob}
\end{equation}
with $p(\mathbf x, t)=\mathbf p(\mathbf x, t)\cdot\mathbf 1 = (p_1(\mathbf x, t), \ldots ,p_m(\mathbf x, t)) \cdot (1, \ldots, 1)^T=\sum\limits_{k=1}^m {p_k}(\mathbf x, t).$
We also define the \textit{time-derivative of the probability vector} $\partial \mathbf p(\mathbf x, t)/\partial t$ as:
\begin{equation*}
\frac{\partial \mathbf p(\mathbf x, t)}{\partial t} \equiv \left( \frac{\partial p_1(\mathbf x, t)}{\partial t} \right., \ldots ,\left. \frac{\partial p_m(\mathbf x, t)}{\partial t} \right),
\end{equation*}
and we have:
\begin{eqnarray*}
\frac{\partial p(\mathbf x, t)}{\partial t} =  \left( \frac{\partial p_1(\mathbf x, t)}{\partial t} \right., \ldots ,\left. \frac{\partial p_m(\mathbf x, t)}{\partial t} \right) \cdot (1, \ldots, 1)^T=\frac{\partial \mathbf p(\mathbf x, t)}{\partial t} \cdot  \mathbf {1} =\sum\limits_{k=1}^m \frac {\partial p_k(\mathbf x, t)}{\partial t}.
\end{eqnarray*}

\subsection{Ordering Microstates, Directional Derivative, and Discrete Divergence}
\label{Ordering Microstates, Directional Derivative, and Discrete Divergence}

\textbf{Ordering Microstates.}
As the microstates are discrete and the stochastic jumps are
dictated by the discrete increments $\{\mathbf s_k\}$ of reactions, we introduce 
\textit{discrete partial derivative} and \textit{discrete divergence} 
to describe effect of specific reactions.
\begin{figure}[thp]
\centering
\includegraphics[scale=1.0]{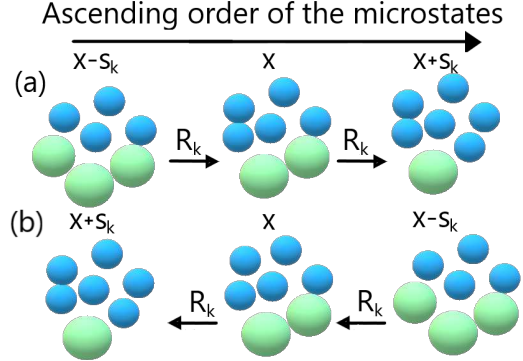}
\caption{ Ordering of microstates: a) when the order of the state preceeding the reaction $R_k$ and the state after the reaction
coincides with the imposed ascending order of microstates, we have $\mathbf x-\mathbf s_k \prec \mathbf x \prec \mathbf x+\mathbf s_k$; 
b) when the order of the state preceding the reaction $R_k$ and the state after the reaction
is in the opposite direction to the ascending order of the microstates, we have $\mathbf x+\mathbf s_k \prec \mathbf x \prec \mathbf x-\mathbf s_k$.}
\label{fig:ord}
\end{figure}

First, we imposed an unambiguous order relationship $ ''\prec''$ over all microstates. 
We impose an ascending order on the microstates $\mathbf x^{0} \prec \mathbf x^{1} \prec \ldots \prec \mathbf x^{\left| \Omega  \right|} $ that is maintained at all time,
such that for each pair of states $\mathbf x^{i} \ne  \mathbf x^{j}$,
either $\mathbf x^{i}  \prec  \mathbf x^{j}$ or
$\mathbf x^{j}  \prec  \mathbf x^{i}$ holds, but not both.
There are many ways to impose such an ordering.
Without loss of generality, we can first use 
the lexicographic order so the microstates are initially sorted by species alphabetically, 
and then by increasing number of molecules of the species. Other ordering schemes are 
also possible.

\textbf{Discrete Partial Derivative.}
We now consider reactional component $p_k(\mathbf x, t)$ of the probability of the state $\mathbf x$ (see Eq.(\ref{eqn:vprob})). For reaction $R_k$,
the only possible change in $\mathbf x$ is determined by its discrete increment of $\mathbf s_k$.

We first consider the case when the state $\mathbf x-\mathbf s_k$ preceding the reaction $R_k$ and the state $\mathbf x$ after the reaction have the order $\mathbf x-\mathbf s_k \prec \mathbf x$. This  also implies $\mathbf x \prec \mathbf x+\mathbf s_k$.
In this case, the direction of the reaction coincides with the direction of the
imposed ordering of the microstates (Figure~\ref{fig:ord}a).
We define the \textit{discrete partial derivative} $
{\Delta p_{k} (\mathbf x, t)/\Delta \mathbf x_k}$ of $p_k(\mathbf x, t)$ over the discrete states in the direction $\mathbf s_k$ of reaction $R_k$
as:
\begin{equation}
\frac {\Delta p_{k}(\mathbf x, t) } {\Delta \mathbf x_k}\equiv {p_{k}(\mathbf x, t) - p_{k}(\mathbf x - \mathbf s_k, t),}
\label{eqn:dder1}
\end{equation}
if $\mathbf x-\mathbf s_k \prec \mathbf x \prec \mathbf x+\mathbf s_k$. 

We now consider the case when $\mathbf x \prec \mathbf x-\mathbf s_k$, namely, 
when the state $\mathbf x - \mathbf s_k$ preceding reaction $R_k$ and the state $\mathbf x$ after $R_k$ are
ordered such that the after-reaction state $\mathbf x$ is placed prior to the before-reaction state $\mathbf x- \mathbf s_k$. This also implies $\mathbf x+\mathbf s_k \prec \mathbf x$  (Figure~\ref{fig:ord}b).
In this case, the discrete partial derivative $
{\Delta p_{k} (\mathbf x, t)/ \Delta \mathbf x_k}$ is defined as: 
\begin{equation}
\frac {\Delta p_{k} (\mathbf x, t)}{\Delta \mathbf x_k} \equiv -{(p_{k}(\mathbf x, t) - p_{k}(\mathbf x + \mathbf s_k, t)),}
\label{eqn:dder2}
\end{equation}
if $\mathbf x + \mathbf s_k \prec \mathbf x \prec \mathbf x - \mathbf s_k$. The negative sign ``--" indicates that the direction of the reaction $R_k$ is opposite to the direction of the imposed order of the states.

\textbf{Discrete Divergence.} We now introduce the \textit{discrete divergence} $\nabla_{d} \cdot \mathbf p (\mathbf x,t) \in \mathbb R$ for the probability vector $\mathbf p(\mathbf x, t)$ 
over the $m$ discrete increments $\left\{ {\mathbf s_k} \right\}$ of the reactions. Applying  Eq.(\ref{eqn:dder1})--(\ref{eqn:dder2}) to each reactional component $p_i (\mathbf x, t)$ of $\mathbf p (\mathbf x, t)$ defined in Eq.(\ref{eqn:vprob}), the discrete divergence $\nabla_{d} \cdot \mathbf p (\mathbf x,t)$  at $\mathbf x$ is the sum of all discrete partial derivatives along the directions of reactions:
\begin{equation}
\nabla_{d} \cdot \mathbf p (\mathbf x, t) \equiv \sum\limits_{k=1}^{m} {\frac {\Delta p_{k} (\mathbf x,t)} {\Delta \mathbf x_k}} .
\label{eqn:div}
\end{equation}

\subsection{Discrete Flux and Velocity at a Fixed Microstate}
\label{Discrete Flux and Velocity at a Fixed Microstate}

\textbf{Single-Reactional Flux.} There are two types of reaction events affecting flux between two states $\mathbf x$ and $\mathbf x + \mathbf s_k$: reactions generating flux flowing from $\mathbf x$ to $\mathbf x + \mathbf s_k$, and reactions generating flux flowing from $\mathbf x + \mathbf s_k$ to $\mathbf x$. The ordering of the microstates enables unique definition of the type of events that the firing of a reaction $R_k$ belongs to. For any two states $\mathbf x$ and $\mathbf x + \mathbf s_k$, only one of the two orderings is possible: we have either $\mathbf x \prec \mathbf x + \mathbf s_k$, or $\mathbf x  + \mathbf s_k \prec \mathbf x$. We define the  \textit{single-reactional flux of probability} $J_k(\mathbf x,t) \in \mathbb{R}$ for reaction $R_k$ at microstate 
$\mathbf x \in \Omega$ as:
\begin{equation}
J_k(\mathbf x,t) \equiv
\left\{ {\begin{array}{*{20}l}
   A_{k} (\mathbf x) p (\mathbf x, t), \, \mathbf x \prec \mathbf x + \mathbf s_k, \\
   A_{k} (\mathbf x-\mathbf s_k) p (\mathbf x - \mathbf s_k, t), \, \mathbf x \prec \mathbf x - \mathbf s_k. \\
 \end{array} } \right.
\label{eqn:flux}
\end{equation}
$J_k(\mathbf x,t)$ depicts the change in $p (\mathbf x,t)$ at the state $\mathbf x$ due to one firing of reaction $R_k$. If $\mathbf x \prec \mathbf x + \mathbf s_k$, $J_k(\mathbf x,t)$ depicts the  outward flux (outflux) of probability due to one firing of reaction $R_k$ at $\mathbf  x$ to bring the system from $\mathbf x $ to $\mathbf x +  \mathbf s_k$. If $\mathbf x \prec \mathbf x -\mathbf s_k$, $J_k(\mathbf x,t)$ depicts the inward flux (influx) of probabability due to one firing of reaction $R_k$ at $\mathbf x -  \mathbf s_k$ to bring the system from $\mathbf x -  \mathbf s_k$ to $\mathbf x$.  For any two states connected by a reaction $R_k$, only one of two orderings is possible as the imposed ordering of the states is unique. Therefore, the single-reactional flux can be applied to all microstates in a self-consistent manner. It also accounts for all reactions,
as $J_k(\mathbf x,t)$ can be defined for every reaction $R_k$. The  \textit{single-reactional $R_k$ velocity} is defined correspondingly  as:
\begin{equation*}
v_k (\mathbf x,t)  \equiv J_k (\mathbf x, t) / p (\mathbf x, t).
\end{equation*}

\textbf{Flux at Boundary States.}  No reactions are possible
if any of the reactant molecules is unavailable, or if its copy number is inadequate.  If $\mathbf x \prec \mathbf x + \mathbf s_k$  (Figure~\ref{fig:ord}a), but  $\mathbf x \in \partial \Omega_k$ ( Eq.(\ref{eqn:bc})), reaction $R_k$ cannot happen, and we have $J_k(\mathbf x,t) = 0$.  If $\mathbf x \prec \mathbf x - \mathbf s_k$ (Figure~\ref{fig:ord}b), but  $\mathbf x  - \mathbf s_k \in \partial \Omega_k$  (Eq.(\ref{eqn:bc})),  reaction $R_k$ cannot happen, and we have $J_k(\mathbf x,t) = 0$. We therefore have the following boundary conditions for $J_k(\mathbf x,t)$:
\begin{equation*}
J_k(\mathbf x,t) \equiv
\left\{ {\begin{array}{*{20}l}
   0, \quad \mathbf x \prec \mathbf x+\mathbf s_k \,\, \textrm{and} \,\, \mathbf x \in \partial \Omega_k\\
  0, \quad \mathbf x \prec \mathbf x-\mathbf s_k \,\, \textrm{and} \,\, \mathbf x-\mathbf s_k \in \partial \Omega_k\\
 \end{array} } \right.
\end{equation*}

\textbf{Discrete Derivative of $J_k$.} Similar to Eq.~(\ref{eqn:dder1}-\ref{eqn:dder2}), the directional derivative of single-reactional flux ${\Delta J_{k} (\mathbf x, t)/ \Delta \mathbf x_k}$ of $J_k(\mathbf x,t)$ along the direction $\mathbf s_k$ of reaction 
$R_k$ is defined as follows:
\begin{equation*}
\frac {\Delta J_{k} (\mathbf x, t)} {\Delta \mathbf x_k} \equiv
\left\{ {\begin{array}{*{20}l}
  \quad A_{k} (\mathbf x) p (\mathbf x, t)-A_{k} (\mathbf x-\mathbf s_k) p (\mathbf x - \mathbf s_k, t), \quad  \quad \quad \quad \quad \quad \,\, \quad \quad \quad \textrm{if} \quad \mathbf x-\mathbf s_k \prec \mathbf x, \\
   -(A_{k} (\mathbf x-\mathbf s_k) p (\mathbf x - \mathbf s_k, t)-A_{k} (\mathbf x \underline{-\mathbf s_k+\mathbf s_k}) p (\mathbf x \underline{- \mathbf s_k+\mathbf s_k}, t)), \quad \textrm{if} \quad \mathbf x \prec \mathbf x-\mathbf s_k. \\
 \end{array} } \right.
\end{equation*}
With simplifications from the trivial identity $\underline{-\mathbf s_k+\mathbf s_k}=0$,
the two expressions of  ${\Delta J_{k} (\mathbf x, t)/ \Delta \mathbf x_k}$ can be combined into one:
\begin{eqnarray}
\frac {\Delta J_{k} (\mathbf x, t)} {\Delta \mathbf x_k}  \equiv A_{k} (\mathbf x)  p (\mathbf x, t)-A_{k} (\mathbf x-\mathbf s_k) p (\mathbf x - \mathbf s_k, t)=- \frac{\partial p_k(\mathbf x, t)}{\partial t}.
\label{eqn:dj}
\end{eqnarray}

\textbf{Total Reactional  Flux, Divergence and Continuity Equation.} We now define the \textit{total reactional flux} or \textit{r-flux} $\mathbf J_r(\mathbf x,t)$, which describes the probability flux at a microstate $\mathbf x$ at time $t$:
\begin{equation}
\mathbf J_r(\mathbf x, t) \equiv ( \, J_{1}(\mathbf x, t), .., \, J_{m}(\mathbf x,t)) \in \mathbb{R}^m.
\label{eqn:je}
\end{equation}
Intuitively, the r-flux $\mathbf J_r(\mathbf x,t)$ is the vector of rate change of the probability mass at $\mathbf x$ in directions of all reactions.
Similar to Eq.~(\ref{eqn:div}), we have the \textit{discrete divergence} of $\mathbf J_{\text{r}}(\mathbf x)$  at microstate $\mathbf x$ :
\begin{equation}
\nabla _{d} \cdot {\mathbf J_r(\mathbf x, t)} \equiv \sum\limits_{k=1}^{m} { \frac {\Delta J_{k} (\mathbf x,t)}{\Delta \mathbf x_k}}   
\label{eqn:ce0}
\end{equation}
From Eq.~(\ref{eqn:dj}) we have:
\begin{eqnarray}
\nabla _{d} \cdot {\mathbf J_r(\mathbf x, t)} = \sum\limits_{k=1}^m \normalsize [ {A_{k} (\mathbf x)  p (\mathbf x,t)} -{A_{k} (\mathbf x-\mathbf s_k)  p (\mathbf x -\mathbf s_k,t)} \normalsize].
\label{eqn:ce1}
\end{eqnarray}

Similar to its continuous version~\cite{shankar2012principles, xu2014energetic} the
discrete continuity equation for the probability mass insists that:
\begin{equation}
\nabla _{d} \cdot {\mathbf J_r(\mathbf x,t)} = - \frac{{\partial p(\mathbf x, t)}}{{\partial t}}.
\label{eqn:coneq}
\end{equation}
From  Eqs.~(\ref{eqn:ce0}), (\ref{eqn:coneq}) and (\ref{eqn:dcme1}), it is clear that r-flux $\mathbf J_r(\mathbf x,t)$ satisfies the continuity equation. The probability mass flows simultaneously along all $m$ directions, with the continuity equation satisfied at all time.

\textbf{Single-Reactional Species Flux and Stoichiometric Projection.} The reactional probability flux $J_k (\mathbf x,t)$ along the direction of reaction $R_k$  defined in Eq.~(\ref{eqn:flux})
can be further decomposed into components of individual species. With the predetermined stoichiometry $\mathbf s_{k} = (s_{k}^{1}, .., \, s_{k}^{n})$, we define the \textit{stoichiometric projection} of  $J_{k} (\mathbf x,t)$ into the component of the $j-$th species $X_j$ 
as:
$$
J^{j}_{k} (\mathbf x,t)  \equiv s_{k}^{j} J_k (\mathbf x,t).
$$

The set of scalar components of all species $\{J^{j}_{k} (\mathbf x,t)\}$ can be used to form a vector $\mathbf J_k (\mathbf x,t) \in \mathbb{R}^n$, which we call \textit{the single-reaction species flux }:
\begin{equation*}
\mathbf J_k (\mathbf x,t) \equiv (J^{1}_{k} (\mathbf x,t), .., J^{n}_{k} (\mathbf x,t))= \mathbf s_{k} J_k(\mathbf x,t) \in \mathbb R^n.
\end{equation*}
\textit{The single-reaction species velocity} of probability is defined correspondingly as $
\mathbf v_k (\mathbf x,t)  \equiv \mathbf J_k (\mathbf x, t) / p (\mathbf x, t).
$

\textbf{Total Species Flux and Velocity.} The \textit{total species flux} or \textit{s-flux} 
$\mathbf J_s (\mathbf x,t)  \in \mathbb{R}^n$ is the sum of all $k$ single-reaction species flux vectors
at a microstate $\mathbf x \in \mathbb{R}^n$:
\begin{equation}
\mathbf J_s (\mathbf x,t)  \equiv \sum\limits_{{\text{k = 1}}}^{\text{m}} \mathbf J_k (\mathbf x, t)=\sum\limits_{{\text{k = 1}}}^{\text{m}} \mathbf s_k J_k (\mathbf x, t) \in \mathbb R^n.
\label{eqn:fl2}
\end{equation}
 The \textit{total species velocity} for probability is defined accordingly as:
\begin{equation}
\mathbf v_s (\mathbf x,t)  = \sum\limits_{{\text{k = 1}}}^{\text{m}} \mathbf J_s (\mathbf x, t) / p (\mathbf x, t).
\label{eqn:v}
\end{equation}
 The s-flux $\mathbf J_s (\mathbf x,t) $ is different from the r-flux $\mathbf J_r(\mathbf x,t)$ defined in Eq.~(\ref{eqn:ce1}). Reaction-centric $\mathbf J_r(\mathbf x,t) \in \mathbb{R}^m$ characterizes the total probability flux at current state in the directions of all reactions, while species-centric $\mathbf J_s (\mathbf x,t) \in \mathbb{R}^n$ sums up the contributions of every reaction to the probability flux at state $\mathbf x$ in the directions of all species.

\subsection{Flux of reversible reaction}
\label{Flux of reversible reaction}

\textbf{Flux of reversible reactions system.} We now 
discuss probability flux in reversible reaction systems that has been previously studied~\cite{zhang2012stochastic, ge2012stochastic},
and how they are related to fluxes formulated here.  
For a pair of the reactions, its directionality needs to be specified upfront,
namely, which reaction is the forward reaction  $R^{+}$, and which is the reversed reaction $R^{-}$:
\begin{eqnarray}
&&R^{+}: c_{1} X_1 + \cdots + c_{n} X_n \stackrel{r^{+}}{\rightarrow} c'_{1} X_1 + \cdots + c'_{n} X_n,\nonumber \\\
&&R^{-}: c'_{1} X_1 + \cdots + c'_{n} X_n 
\stackrel{r^{-}}{\rightarrow} c_{1} X_1 + \cdots + c_{n} X_n. \nonumber
\end{eqnarray}
Let $\mathbf s=( \, c'_{1} - c_{1},\ldots ,  c'_{n}-c_{n}) \,$ be the stoichiometry of reaction $R^{\text{+}}$, $- \mathbf s$ the stoichiometry of reaction $R^{-}$.
The flux $J$ described in~\cite{zhang2012stochastic, ge2012stochastic} is the net flux between $\mathbf x$ and $\mathbf x+ \mathbf s$. It is specified as the difference between the forward flux at $\mathbf x$ $J^{+}(\mathbf x, t) = r^{+} \prod\limits_{l = 1}^n {{x_l }  \choose {c_l }}  p (\mathbf x, t)$ generated by the forward reaction  $R^{\text{+}}$ and the reverse flux at $\mathbf x + \mathbf s$ $J^{\text{-}} (\mathbf x+ \mathbf s, t) = r^{ - }  \prod\limits_{l = 1}^n  {{x_l + s_{l}}  \choose {c'_l }} p (\mathbf x + \mathbf s,t)$ generated by the reverse reaction $R^{-}$, both connecting $\mathbf x$ and $\mathbf x + \mathbf s$~\cite{zhang2012stochastic, ge2012stochastic}: 
\begin{eqnarray}
J(\mathbf x, t) =  r^{+} \prod\limits_{l = 1}^n {{x_l }  \choose {c_l }}  p (\mathbf x, t) -  r^{ - }  \prod\limits_{l = 1}^n  {{x_l + s_{l}}  \choose {c'_l }} p (\mathbf x + \mathbf s,t).
\label{eqn:fplus}
\end{eqnarray}
\textbf{Conversion between single-reactional species flux and flux in a pair of reversible reaction system.} The flux $J(\mathbf x, t)$ for a pair of reversible reactions above can be related to the s-flux $\mathbf J_s (\mathbf x, t)$ of Eq.~(\ref{eqn:fl2}) by examining the projection of the $J(\mathbf x, t)$ in
Eq.~(\ref{eqn:fplus}) to individual species.
Specifically, with the stoichiometry  $\mathbf s$, the projection of the flux of Eq.~(\ref{eqn:fplus})  to the component of the $j-$th species $X_j$  
is:
\begin{eqnarray}
\mathbf J(\mathbf x,t) =  \mathbf {s} J (\mathbf x,t)=\mathbf s  r^{+} \prod\limits_{l = 1}^n {{x_l }  \choose {c_l }}  p (\mathbf x, t)
- \mathbf s  r^{ - }  \prod\limits_{l = 1}^n  {{x_l + s_l}  \choose {c'_l }} p (\mathbf x+ \mathbf s,t) \in \mathbb{R}^n.
\label{eqn:fluxQ}
\end{eqnarray}
When the direction of the forward reaction $R^{+}$ coincides with the ascending order of the states,  one firing of $R^{+}$ with the stoichiometry
vector $\mathbf s$  at the state $\mathbf x$ brings the system to the state $\mathbf x + \mathbf s$ in the direction of the ascending order.
From Eq.~(\ref{eqn:fl2}), the s-flux $\mathbf{J}_{s}(\mathbf{x},t)$ for $(R^{+}, R^{-})$ is
$\mathbf J_s (\mathbf x, t)=\mathbf s  r^{\text{ + }} \prod\limits_{l = 1}^n {{x_l }  \choose {c_l }}  p (\mathbf x, t) - \mathbf s r^{\text{ - }}  \prod\limits_{l = 1}^n  {{x_l+s_l }  \choose {c'_l }} p (\mathbf x + \mathbf s, t).$  
In this case, the projection of the reversible reaction flux by
Eq.~(\ref{eqn:fluxQ})  is identical to the s-flux by Eq.~(\ref{eqn:fl2}) at the state $\mathbf x$.

When the direction of the forward reaction $R^{+}$ 
is opposite to the ascending order of the states,  one firing of $R^{-}$with the stoichiometry
vector $- \mathbf s$  at the state $\mathbf x + \mathbf s $ brings the system to the state $\mathbf x$ in the direction of the ascending order.
From Eq.~(\ref{eqn:fl2}), the s-flux $\mathbf{J}_{s}(\mathbf{x}+\mathbf s, t)$ for $(R^{+}, R^{-})$ is
$\mathbf J_s
(\mathbf x + \mathbf s, t) = \mathbf {s}  r^{\text{ + }} \prod\limits_{l = 1}^n {{x_l }  \choose {c_l }}  p (\mathbf x, t)  -\mathbf {s} r^{\text{ - }}  \prod\limits_{l = 1}^n  {{x_l +s_l}  \choose {c'_l }} p (\mathbf x + \mathbf s, t)$ .
In this case, the projection of the reversible reaction flux by
Eq.~(\ref{eqn:fluxQ})  is identical to s-flux by Eq.~(\ref{eqn:fl2}) at the state $\mathbf x + \mathbf s$.

\section{Results}

\label{Results}

Below we illustrate how time-evolving and steady-state flux and velocity
fields of the probability mass can be computed for three model systems,
namely, the birth-death process, the bistable Schl\"ogl model, and the oscillating Schnakenberg system.  
The underlying
discrete Chemical Master Equation (dCME) (Eq.(\ref{eqn:dcme1})) of these models are solved using the recently developed ACME method~\cite{cao2016state, cao2016accurate}.
The resulting exact probability landscapes of these models are used to compute the flux and the velocity fields. 
 
\subsection{The Birth and Death Process}
The birth-death process is a simple, but ubiquitous process of 
the synthesis and degradation of molecule of a single specie~\cite{allen2010introduction,cao2016state}. 
The reaction schemes and rate 
constants examined in this study are specified as follows: 
\begin{eqnarray*}
&&R_1: \quad \emptyset \stackrel{r_1}{\rightarrow} X, \quad r_1 = 1, \nonumber\\
&&R_2: \quad X \stackrel{r_2}{\rightarrow} \emptyset, \quad r_2 = 0.025. 
\end{eqnarray*}
Below we use $k$ as the index of the two reactions.

\textit{Ordering Microstates.} The microstate in this system is defined by the copy number $x$ of the molecular specie $X$.  We order the microstates in the direction of increasing copy numbers of $x$, namely,  $(x=0)\, \prec \,   (x=1) \,  \prec \,  (x=2) \cdots$.

\textit{Discrete Increment and Reaction Direction.} Reaction $R_1$ brings the system from the state $x$ to the state $x+1$, in the direction of increasing order of the microstates. Its discrete increment is $s_1=1$. Reaction $R_2$ brings the system from the state $x$ to the state $x-1$, in the direction of decreasing order of the microstates. Its discrete increment is therefore $s_2=-1$.  

\textit{Discrete Chemical Master Equation.} Following Eq.(\ref{eqn:dcme1}), the discrete Chemical Master Equation for this system can be written as:
\begin{eqnarray}
\partial p (x,t)/ \partial t = r_1 p(x,t)-r_1 p(x-1,t)-r_2 (x+1)  \times p(x+1,t) + r_2  x p(x,t).
\label{eqn:bd_dcme}
\end{eqnarray}

\textit{ Single-Reactional Flux, Velocity and Boundary Conditions.}
The  single-reactional flux $J_k (x,t) \in  \mathbb{R}$ can be written as: 
\begin{equation}
J_1(x,t) = r_1 p(x,t), \,\, J_2(x,t) = r_2 (x+1) p(x+1,t).
\label{eqn:f_b}
\end{equation}
Here $x=0, 1,...$. No special boundary conditions are required for this system, as  $J_1(x,t)$ and $J_2(x,t)$ at the boundary $x=0$ take the values specified by Eq.~(\ref{eqn:f_b}).
The single-reactional velocity $v_k (x,t) \in  \mathbb{R}$ can be written as:
$v_1(x,t) = J_1(x,t)/p(x,t)\quad$ and  $ \quad  v_2(x,t) = J_2(x,t)/p(x,t).$

\textit{Discrete Partial Derivative.} 
The imposed ordering of the microstates implies
$x \prec x + s_1$, as $s_1=1$ and $x \prec x + 1$. By Eq.~(\ref{eqn:dder1}), the derivative $ {\Delta J_{1} (x, t)} /{\Delta x_1} $ of the single-reactional flux function  $J_1$  is:
\begin{eqnarray}
\frac {\Delta J_{1} (x, t)} {\Delta  x_1} =  J_{1} (x, t) - J_{1} (x- s_1, t) 
= &r_1 p(x,t)-r_1 p(x-1,t). \nonumber
\end{eqnarray}
The imposed ordering of the microstates also has
$x \prec x - s_2$, as $s_2=-1$ and $x \prec x + 1$.
By Eq.~(\ref{eqn:dder2}), the derivative $ {\Delta J_{2} (x, t)}/ {\Delta x_2} $ of the single-reactional flux function  $J_2$ is:
\begin{eqnarray}
\frac {\Delta J_{2} (x, t)} {\Delta x_2} = -(J_{2} (x, t) - J_{2} (x + s_2, t))= -(r_2 (x+1) p(x+1,t) - r_2 (x) p(x,t)).\nonumber
\end{eqnarray}

\textit{Total Reactional Flux, Discrete Divergence, and Continuity Equation.}  Following  Eq.~(\ref{eqn:je}), the total reactional flux
$\mathbf J_r(x,t) \in  \mathbb{R}^{2}$ is:
\begin{eqnarray}
\mathbf J_r(x,t)= (J_1(x,t),\,J_2(x,t)) = (r_1 p(x,t),\,\, r_2 (x+1) p(x+1,t)).\nonumber 
\end{eqnarray}
The total reactional velocity $\mathbf v_r( x,t) \in  \mathbb{R}^{2}$ is:
$
\mathbf v_r(x,t) = \mathbf J_r(x,t)/ p(x,t).
$

Following  Eq.~(\ref{eqn:div}), the discrete divergence $\nabla_{d} \cdot \mathbf J_r (x, t)$ of $\mathbf J_r (x, t) \in \mathbb{R}^{2}$  over the discrete increments $s_1$ and $s_2$ can be written as:
\begin{eqnarray}
\nabla_{d} \cdot \mathbf J_r (x, t) \equiv && \sum\limits_{k=1}^{2} {\frac {\Delta J_{k} (x,t)} {\Delta x_k}} = r_1 p(x,t)-r_1 p(x-1,t)\nonumber \\
&& -r_2 (x+1) p(x+1,t) + r_2 (x) p(x,t).
\label{eqn:coneq1}
\end{eqnarray}

Here the r-flux $J_r (x, t)$ indeed satisfies the continuity equation, as we have 
$\nabla_{d} \cdot \mathbf J_r (x, t)=-\partial p(x,t)/\partial t$ from  Eqs.~(\ref{eqn:coneq}), (\ref{eqn:bd_dcme}), and (\ref{eqn:coneq1}).

\textit{Stoichiometry projection and single-reactional species flux.} Since there is only one specie in this system, the stoichiometry projection of $J_k(x,t)$ to the specie $X$ equals to the single-reactional species flux $\mathbf J_k(x,t)\in \mathbb{R}$, which can be written as: 
$$\mathbf J_1(x,t) = r_1 p(x,t) \quad \text{and} \quad \mathbf J_2(x,t) = - r_2 (x+1) p(x+1,t).$$
The single-reactional species velocity $\mathbf v_k(x,t) \in \mathbb{R}$ can be written as follows:
$
\mathbf v_1(x,t) = J_1(x,t)/p(x,t)\quad$ and $ \quad \mathbf v_2(x,t) = J_2(x,t)/p(x,t).
$

\textit{Total Species Flux and Velocity.} Following Eqs.~(\ref{eqn:fl2})--(\ref{eqn:v}), the  s-flux $J_s (x,t) $ and the total velocity $v_s (x,t) $ are:
$$J_s(x,t)= r_1 p(x,t)-r_2 (x+1) p(x+1,t),$$
$$v_s(x,t)=J_s(x,t)/p(x,t).$$
When $J_s(x,t)>0$ and $v_s(x,t)>0$, the probability mass moves in the direction of increasing 
copy number of $X$. This is the direction of the ascending order of microstates we imposed. When $J_s(x,t)<0$ and $v_s(x,t)<0$,
the probability mass moves in the direction of the decreasing 
copy number of $X$. We will further use just simple flux instead of s-flux.

\textit{Overall Behavior of the Birth and Death System.}  We examine the behavior of the birth and death process under the initial conditions $\left. {p(x = 0)} \right|_{t = 0}=1 $ (Figure~\ref{fig:n_bd}a, backside)
and that of the uniform distribution (Figure~\ref{fig:n_bd}d, backside). 

For the initial
condition of $\left. {p(x = 0)} \right|_{t = 0}=1 $, the probability landscape changes from that with a peak at $x = 0$ to 
that with a peak at $x=40$ (Figure~\ref{fig:n_bd}a). 
Figure~\ref{fig:n_bd}b shows the heatmap of the flux $J_s(x,t)$, and Figure~\ref{fig:n_bd}c the heatmap of the velocity $v_s(x,t)$. Yellow and red areas represent locations where the probability moves in the positive direction, while white areas represents locations where the flux and velocity both are close to be zero. The flux and velocity of probability mass (Figure~\ref{fig:n_bd}b-- ~\ref{fig:n_bd}c)
are positive at all time, indicating that the probability mass is  moving only in the direction
of increasing copy number of $x$. 
Moreover,  when the probability is non-zero, the probability velocity remains constant at any fixed time $t$ across different microstates.
 The blue line in Figure~\ref{fig:n_bd}b--~\ref{fig:n_bd}c corresponds to the peak of the system, that changes its location from $x=0$ to $x=40$.

For the initial condition of the uniform distribution,  the probability landscape changes from the constant line to that with a peak at $x=40$ (Figure~\ref{fig:n_bd}d).
Figure~\ref{fig:n_bd}e shows the heatmap of the flux $J_s(x,t)$, and Figure~\ref{fig:n_bd}f the heatmap of the velocity $v_s(x,t)$. Blue areas represent locations where the probability mass moves in the negative direction, yellow and red areas represent locations where the probability moves in the positive direction, while white areas represents locations where the flux and velocity both are equal to zero.
Specifically, when $x<40$, we have $J_s(x,t)>0$ and $v_s(x,t)>0$, namely, the probability mass moves in the direction of increasing copy number of $x$. In contrast, when $x>40$, we have $J_s(x,t)<0$ and $v_s(x,t)<0$, indicating that the probability mass moves in the direction of decreasing copy number of $x$.
When $x=40$, we have $J_s(x,t)=0$ and $v_s(x,t)=0$.
Furthermore, the probability velocity at a specific time $t$ 
is different for different microstates, with the highest velocities located at the boundary of $x=0$.
 The blue line in Figure~\ref{fig:n_bd}e--~\ref{fig:n_bd}f $x=40$ corresponds to the peak of the system, which appears starting at about $t=5$.

To solve this problem using the ACME method, we introduced the buffer of capacity $x=92$. At the state $x=92$ when the buffer is exhausted, no synthesis reaction can occur. Therefore, the flux at the boundary $x=92$ is set to zero.

Our birth and death system eventually reaches to a steady state. 
As expected, the same steady state probability distribution 
is reached from both initial conditions (shown in different scale in
Figure~\ref{fig:n_bd}a  and~\ref{fig:n_bd}d). At the steady state, the probability landscape has a peak at 
$x=40$. Both the velocity $v_s(x,t)$ and the flux $J_s(x,t)$ converge to zero at steady state. 
\begin{figure*}[thp]
\centering
\includegraphics[scale=1.00]{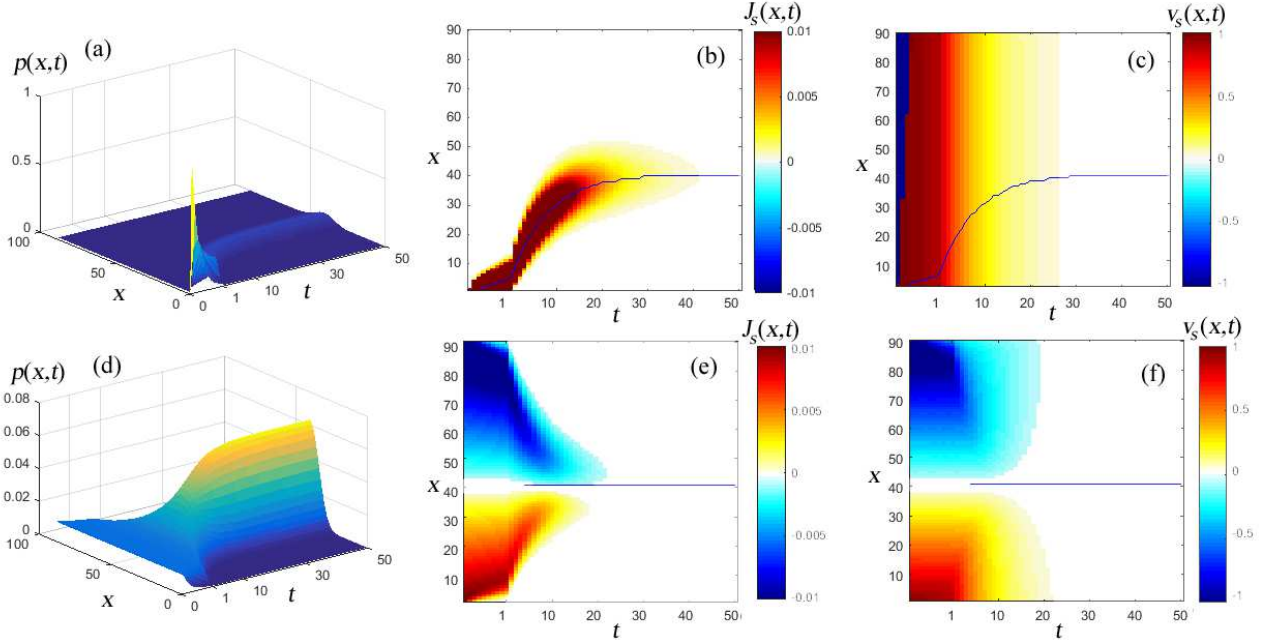}
\caption{ The time-evolving probability landscape, flux and velocity of
the probability mass of the birth and death system
starting from the initial conditions of $\left. {p(x = 0)} \right|_{t = 0}=1 $ (a--c) and from  
the initial conditions of 
the uniform distribution (d--f). 
a) and d):  the probability landscape in $p(x,t)$;  
b) and e): the corresponding value of flux $J_s(x,t)$; 
c) and f): the value of velocity $v_s(x,t)$.
}
\label{fig:n_bd}
\end{figure*}

\subsection{Bistable Schl\"ogl model}
The Schl\"ogl model is a one-dimensional bistable system consisting of an auto-catalytic network involving one molecular specie $X$ 
and four reactions~\cite{schlogl1972chemical}. It is a canonical model for studying bistability and state-switching~\cite{vellela2009stochastic,  cao2013adaptively}. The reaction schemes and kinetic constants examined in this study are specified as follows:
\begin{equation}
\begin{aligned}
&R_1: \quad A + 2X\stackrel{k_1}{\rightarrow} 3X, \quad k_1 = 6; \\
&R_2: \quad 3X\stackrel{k_2}{\rightarrow} A + 2X, \quad k_2 = 3.6; \\
&R_3: \quad B \stackrel{k_3}{\rightarrow} X, \quad k_3 = 0.25;\\
&R_4: \quad X\stackrel{k_4}{\rightarrow} B, \quad k_4 = 2.95.
\label{eqn:s_react}
\end{aligned}
\end{equation}
Here $A$ and $B$ have constant  concentrations $a$ and $b$,
which are set to $a = 1$ and $b = 2$, respectively. We set the volume of the system to $V=25$~\cite{schlogl1972chemical}. The rate of reactions are specified as $r_1=k_1/V$,  $r_2=k_2/V^2$,  $r_3=k_3V$,  $r_4=k_4$.

\textit{Ordering Microstates.} We define the microstates of this system using the copy number $x$ of the molecular specie $X$.  We order the microstates in the direction of increasing copy numbers of $X$, namely, $(x=0) \, \prec  \, (x=1)  \, \prec  \, (x=2) \cdots$.

\textit{Discrete Increment and Reaction Direction.} Reactions $R_1$ and $R_3$ bring the system from the state $x$ to the state $x+1$, in the direction of increasing order of the microstates. Their discrete increments $s_1$ and $s_3$ are $s_1=1$ and $s_3=1$. Reactions $R_2$ and $R_4$ bring the system from the state $\mathbf x$ to the state $x-1$, in the direction of decreasing order of the microstates. Their discrete increments $s_2$ and $s_4$ are therefore $s_2=-1$ and $s_4=-1$.  

\textit{Discrete Chemical Master Equation.} Following Eq.(\ref{eqn:dcme1}), the discrete Chemical Master Equation for this system can be written as:
\begin{equation}
\begin{aligned}
\frac {\partial p (x,t)}{\partial t}  = & r_1 a \frac {(x-1) (x-2)} {2} p(x-1,t)   + r_2 \frac {(x+1) x(x-1)} {6} p(x+1,t)  + r_3 b p(x-1,t)+ r_4 (x+1) \\
& \times p(x+1,t) - r_1 a \frac {x(x-1)} {2} p(x,t) - r_2 \frac {x(x-1) (x-2)} {6} p(x,t)  - r_3 b P(x,t) - r_4 x p(x,t). 
\label{eqn:s_dcme11}
\end{aligned}
\end{equation}
We compute the probability landscape $p(x,t)$ underlying Eq.(\ref{eqn:s_dcme11}) using the ACME method~\cite{cao2016state, cao2016accurate}.

\textit{ Single-Reactional Flux, Velocity and Boundary Conditions.}
Following Eq. (\ref{eqn:flux}),  the single-reactional flux $J_k(x,t) \in  \mathbb{R}$ can be written as: 
\begin{eqnarray}
&& J_1(x,t) = r_1 a \frac {x(x-1)} {2} p(x,t), J_2(x,t) = r_2 \frac {(x+1) x(x-1)} {6} p(x+1,t), \nonumber \\
&& J_3(x,t) = r_3 b p(x,t), 
 J_4(x,t) = r_4 (x+1) p(x+1,t). \nonumber 
\end{eqnarray}
We have the single-reactional fluxes  $J_{1} (x,t)  = 0$ and $J_{2} (x,t) = 0$ on the boundary with either $x=0$ or $x=1$, where reactions $R_1$ and $R_2$ cannot happen.
The single-reactional fluxes $J_{3} (x,t) $ and $J_{4} (x,t)$  are as given above and do not vanish at the boundaries.

The single-reactional velocity $v_k \in  \mathbb{R}$ can be written as: $v_k(x,t) = J_k(x,t)/p(x,t)$, with $k=1,\ldots,4$.

\textit{Discrete Partial Derivative.}
The imposed ordering of the microstates has $x \prec x + 1$, therefore,
$x \prec x + s_1$ , $x \prec x -  s_2$, $ x \prec x + s_3$, and $x \prec x -  s_4$, as $s_1=1$, $s_2=-1$, $s_3=1$, and $s_4=-1$.
According to Eqs.~(\ref{eqn:dder1}) -- (\ref{eqn:dder2}), the derivatives $ {\Delta J_{k} (x, t)} /{\Delta x_k} $ of the single-reactional fluxes $\{J_k\}$ are:
\begin{eqnarray}
&&\frac {\Delta J_{1} (x, t)} {\Delta x_1} = J_{1} (x, t) - J_{1} (x - s_1, t) 
=r_1 a \frac {x (x -1)} {2} p(x,t)- r_1 a \frac {(x-1) (x -2)} {2} p(x-1,t), \nonumber \\
&&\frac {\Delta J_{2} (x, t)} {\Delta x_2}  = -(J_{2} (x, t) - J_{2} (x + s_2, t)) =- (r_2 \frac {(x+1) x (x-1)} {6} p(x+1,t) - r_2 \frac {(x + 2) (x + 1) x} {6} p(x,t)),\nonumber \\
&&\frac {\Delta J_{3} (x, t)} {\Delta x_3} = J_{3} (x, t) - J_{3} (x - s_3, t)= - (r_3 b p(x,t) - r_3 b p(x - 1,t)),\nonumber \\
&&\frac {\Delta J_{4} (x, t)} {\Delta x_4} = -(J_{4} (x, t) - J_{4} (x + s_4, t)) = - (r_4 (x +1) p(x+1,t) - r_4 x p(x,t)). \nonumber 
\end{eqnarray}
\textit{Total Reactional Flux and Velocity, Discrete Divergence, and Continuity Equation.} Following Eq.~(\ref{eqn:je}),  the total reactional flux $\mathbf J_r(x,t) \in  \mathbb{R}^{4}$ is:
\begin{eqnarray}
\mathbf J_r(x,t)&&=(J_1(x,t), J_2(x,t), J_3(x,t), J_4(x,t)) \nonumber\\
&&=\normalsize ( r_1 a \frac {x (x -1)} {2} p(x,t), \,\, r_2  \frac {(x+1) x (x-1)} {6}  p(x+1,t),\,\,  r_3 b p(x,t), \,\,  r_4 (x +1) p(x+1,t) \normalsize ).\nonumber 
\end{eqnarray}
The total reactional velocity $\mathbf v_r(x,t) \in  \mathbb{R}^{4}$ is: $
\mathbf v_r(x,t) = \mathbf J_r(x,t)/ p(x,t).$

The discrete divergence $\nabla_{d} \cdot \mathbf J_r (x, t)$ of $\mathbf J_r(x, t) \in \mathbb{R}^{4}$ 
over the discrete increments $s_1$, $s_2$, $s_3$, and $s_4$ can be written as:
\begin{equation}
\begin{aligned}
\nabla_{d} \cdot \mathbf J_r (x, t) &=  \sum\limits_{k=1}^{4} {\frac {\Delta J_{k} (x,t)} {\Delta x_k}}=-\frac {(x-1) (x -2)} {2} r_1 a  p(x-1,t)\\
&+r_1 a \frac {x (x -1)}{2} p(x,t)- r_2  \frac {(x+1) x (x-1)} {6} p(x+1,t) \\
&+ r_2 \frac {x (x-1) (x-2)} {6} p(x,t)- r_3 b p(x-1,t) + r_3 b p(x,t) \\
&- r_4 (x +1) p(x+1,t) + r_4 x p(x,t). 
\label{eqn:coneq2}
\end{aligned}
\end{equation}
The flux  $\mathbf J_{\text{ R}} (x, t)$ indeed satisfies the continuity equation, as we have:
$\nabla_{d} \cdot \mathbf J_r (x, t)=-\partial p(x,t)/\partial t$ from Eqs.~(\ref{eqn:coneq}),  (\ref{eqn:s_dcme11}), and (\ref{eqn:coneq2}).

\textit{Stoichiometry projection and single-reactional species flux.} Since there is only one specie $x$ in this system, the stoichiometry projection of single-reactional flux $ J_k(x,t)$ to $x$ equals to the single-reactional species flux $\mathbf J_k(x,t) \in \mathbb{R}$, which can be written as: 
\begin{eqnarray}
&& \mathbf J_1(x,t) = r_1 a \frac {x (x -1) } {2} p(x,t), \nonumber \\
&& \mathbf J_2 (x,t) =- r_2 \frac {(x+1) x (x-1)} {6} p(x+1,t), \nonumber \\
&& \mathbf J_3(x,t) = r_3 b p(x,t), \nonumber \\
&&  \mathbf  J_4(x,t) =  -r_4 (x +1) p(x+1,t). \nonumber 
\end{eqnarray}
The single-reactional species velocities $\mathbf v_k \in \mathbb{R}$ is $\mathbf v_k(x,t) = \mathbf J_k(x,t)/p(x,t)$, with $k=1,\ldots,4$. 

\textit{Total Species Flux and Velocity.} Following Eqs.~(\ref{eqn:fl2})--(\ref{eqn:v}), the total species flux $\mathbf J_s(x,t)$ and velocity $\mathbf v_s(x,t)$ for the four reactions are :
\begin{eqnarray}
J_s(x,t)= r_1 a \frac {x (x-1)}{2} p(x,t) - r_2 \frac {(x+1) x (x-1)} {6} p(x+1,t) + r_3 b p(x,t) - r_4 (x +1) p(x+1,t), \nonumber
\end{eqnarray}
and $v_s (x,t)=J_s(x,t)/p(x,t).$

\textit{Overall Behavior of the Schl\"{o}gl System.} 
For the set of parameter values used in Eqs.~(\ref{eqn:s_react}), Schl\"{o}gl model is bistable. It has two peaks at $x=4$  and $x=92$.
In order to study how switching between the two peaks occur, we examine the behavior of the model under the initial conditions of $\left. {p(x = 4)} \right|_{t = 0}=1 $ (Figure~\ref{fig:sc}a)
and the initial condition of $\left. {p(x = 92)} \right|_{t = 0}=1 $ (Figure~\ref{fig:sc}d).

For the initial distribution of $\left. {p(x = 4)} \right|_{t = 0}=1 $, the probability landscape changes from that with a single peak at $x = 4$ to 
that with two maximum peaks at $x=4$ and $x=92$  (Figure~\ref{fig:sc}a). 
Figure~\ref{fig:sc}b shows the heatmap of the flux $J_s(x,t)$, and Figure~\ref{fig:sc}c the heatmap of the velocity $v_s(x,t)$. Yellow and red areas represent locations where the probability moves in the positive direction, while white areas represents locations where the flux and velocity both are close to be zero.
 The lower blue lines in Figure~\ref{fig:sc}b--~\ref{fig:sc}c correspond to the peak at $x=4$. They are straight lines as the location of the peak does not change over time. Another blue line starts to appear at $x=92$ at about $t=3$ and corresponds to the second peak. At the same time, at around $t=3$, we observe
the appearance of a minimum of the probability landscape (red line),  separating the two maximum peaks.
We have $J_s(x,t)> 0$ and $v_s(x,t)>0$, indicating that the probability moves in the direction of increasing copy number of molecules (Figure~\ref{fig:sc}b--~\ref{fig:sc}c) in the majority of the states. In the white region, we have  $J_s(x,t)= 0$ and $v_s(x,t)=0$.

\begin{figure*}[thp]
\centering
\includegraphics[scale=1.00]{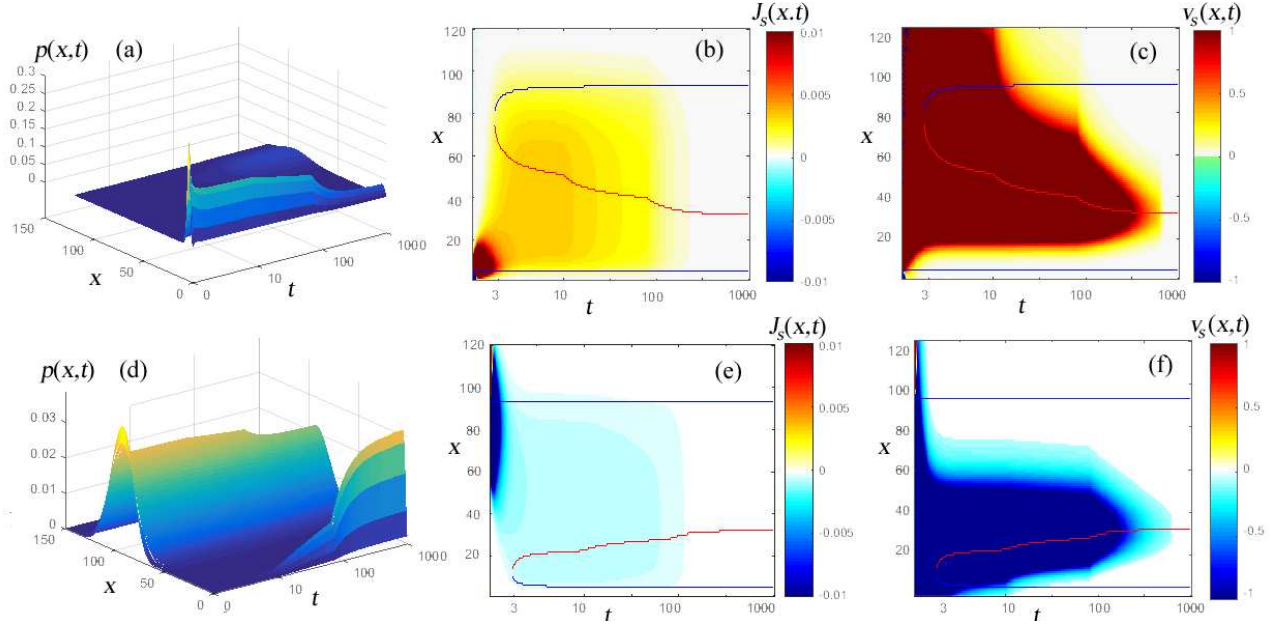}
\caption{
The time-evolving probability landscape, flux and velocity of
the probability mass in the Schl\"{o}gl system
starting from the initial conditions of $\left. {p(x = 4)} \right|_{t = 0}=1 $ (a--c) and from  the initial conditions of $\left. {p(x = 92)} \right|_{t = 0}=1$ (d--f). 
a) and d):  the probability landscape in $p(x,t)$ ;
b) and e): the corresponding value of flux in $J_s(x,t)$; 
c) and f): the value of velocity $v_s(x,t)$.
}
\label{fig:sc}
\end{figure*}

For the first initial condition of  $\left. {p(x = 92)} \right|_{t = 0}=1 $, the probability landscape changes from that with a single peak at $x = 92$ to 
that of two peaks at $x=92$ and $x=4$ (Figure~\ref{fig:sc}d).
Figure~\ref{fig:sc}e shows the heatmap of the flux $J_s(x,t)$, and Figure~\ref{fig:sc}f the heatmap of the velocity $v_s(x,t)$. Blue areas represent locations where the probability mass moves in the negative direction, while white areas represents locations where the flux and velocity both are equal to zero. The top blue lines in Figure~\ref{fig:sc}e--~\ref{fig:sc}f correspond to the peak at $x=92$. These are straight lines as the location of this peak does not change over time. 
Another blue line starting to appear at $x=4$ at around  $t=3$ and corresponds to the second peak. 
At around  $t=3$, we also observe
the appearance of a minimum on the probability landscape (red line) separating the two maximum peaks.
In the blue region, we have  $J_s(x,t)<0$ and $v_s(x,t)<0$,  and the probability moves in the direction of increasing copy number of molecules
(Figure~\ref{fig:sc}e--~\ref{fig:sc}f) in the majority of states. In the white region, we have  $J_s(x,t)= 0$ and $v_s(x,t)=0$.

In both cases (Figure~\ref{fig:sc}), the second peak appears after about $t=3$. We also observe that the absolute values of the flux driving the system from the system with one peak at $x=4$ to the emergence of the second peak at $x=92$, and from the system with one peak at $x=92$ to the emergence of the second peak at $x=4$ are of the same scale.

The Schl\"{o}gl process eventually reaches to a steady state. 
As expected, the same steady state probability distribution 
is reached from both initial conditions. At the steady state, the probability landscape has two peaks at 
$x=4$ and $x=92$. Both the velocity $v_s(x,t)$ and the flux $J_s(x,t)$ converge to zero at the steady state. 

\subsection{Schnakenberg Model}

The Schnakenberg model is a simple chemical reaction system originally constructed to study the behavior of limit cycle~\cite{schnakenberg1976network}.
It provides an important model for analyzing oscillating behavior in reaction systems~\cite{qian2002concentration,  cao2010nonlinear, xu2012energy}.
The reaction scheme and rate constants examined in this study are specified as follows:
\begin{eqnarray*}
&&R_1: \quad A \stackrel{k_1}{\rightarrow} X_1, \quad k_1 = 1;  \nonumber \\
&&R_2: \quad  X_1 \stackrel{k_2}{\rightarrow} \emptyset, \quad  k_2 = 1; \nonumber \\
&&R_3: \quad  B \stackrel{k_3}{\rightarrow} X_2, \quad  k_3 = 1; \quad  \nonumber \\
&&R_4: \quad  X_2 \stackrel{k_4}{\rightarrow} \emptyset, \quad k_4 = 10^{-2};  \\
&&R_5: \quad  2X_1 + X_2 \stackrel{k_5}{\rightarrow} \, 3X_1, \quad k_5 = 1;\nonumber\\
&&R_6: \quad  3X_1 \stackrel{k_6}{\rightarrow} \quad 2X_1+X_2, \quad  k_6 = 10^{-2}. \nonumber
\end{eqnarray*}
Here $X_1$ and $X_2$ are molecular species whose copy numbers $x_1$ and $x_2$ oscillate, $A$ and $B$ are reactants of fixed copy numbers of $a$ and $b$, respectively. 
The volume of the system $V$ is set to $V=10^{-2}$ ~\cite{schnakenberg1976network}.
The rate of reactions are specified as $r_1=k_1$,  $r_2=k_2$,  $r_3=k_3$,  $r_4=k_4$, $r_5=k_5/V^{2}$,  $r_6=k_6/V^{2}$.

\textit{Ordering Microstates.} The microstate $\mathbf x=(x_1,x_2)$  in this system is defined by the ordered pair of copy numbers $x_1$ and $x_2$ of the molecular species $X_1$ and $X_2$. We impose the ascending order of the microstates first in the direction of the increasing copies of $X_1$. At fixed value of 
$X_1$, we then sort the states in the order of increasing copy number of $X_2$. We therefore have 
$(x_1=0, x_2=0) \prec (x_1=0, x_2=1) \prec (x_1=0,  x_2=2)  \prec \cdots  \prec(x_1=1, x_2=0) \prec (x_1=1, x_2=1) \cdots $. 

\textit{Discrete Increment and Reaction Direction.}  The discrete increments $\mathbf s_1$, $\mathbf  s_3$, and $\mathbf  s_5$ of reactions $R_1$, $R_3$, and $R_5$ that bring the system in the direction of increasing order of the microstates and the discrete increments $\mathbf s_2$, $\mathbf  s_4$, and $\mathbf  s_6$ of reactions $R_2$, $R_4$, and $R_6$ that bring the system in the direction of the decreasing order of the microstates are listed in Table~\ref{table:1}.
\begin{table*}
\caption{Schnakenberg system reactions stoichiometry}
\begin{tabular}{|c || c c c | c c c|} 
 \hline
Reactions & $R_1$ & $R_3$ & $R_5$ & $R_2$ & $R_4$ & $R_6$ \\ 
 \hline
 Discrete Increments & $\mathbf s_1=(1,0)$ &  $\mathbf s_3=(0,1)$ & $\mathbf s_5=(1,-1)$ & $\mathbf s_2=(-1,0)$ & $\mathbf s_4=(0,-1)$ & $\mathbf s_6=(-1,1)$\\ 
  \hline
\end{tabular}
\label{table:1}
\end{table*}

\textit{Discrete Chemical Master Equation.} Following Eq.(\ref{eqn:dcme1}), the discrete Chemical Master Equation for the system can be written as:
\begin{equation}
\begin{aligned}
\frac {\partial p (\mathbf x,t)}{\partial t}=&-r_1 a p(x_1,x_2,t) + r_1 a p(x_1-1,x_2,t)  -r_2 x_1 p(x_1,x_2,t) + r_2 (x_1+1) p(x_1+1,x_2,t)\\
& - r_3 b p(x_1,x_2,t)+ r_3 b p(x_1,x_2-1,t)  + r_4 (x_2+1) p(x_1,x_2+1,t)- r_4 x_2 p(x_1,x_2,t) \\
&+ r_5 \frac {(x_1-1) (x_1 -2) x_2} {2} p(x_1-1,x_2+1,t) -r_5 \frac {x_1 (x_1 -1) x_2} {2} p(x_1,x_2,t) \\ 
& +r_6 \frac {(x_1-1) x_1 (x_1+1)} {6} p(x_1+1,x_2-1,t) -r_6 \frac {x_1 (x_1 -1) (x_1-2)} {6} p(x_1,x_2,t). 
\label{eqn:schn_dcme11}
\end{aligned}
\end{equation}
We compute the probability landscape $p(\mathbf x,t)$ underlying Eq.(\ref{eqn:s_dcme11}) using the ACME method~\cite{cao2016state, cao2016accurate}.

\textit{Single-Reactional Flux, Velocity and Boundary Conditions.}
The single-reactional flux $J_k (\mathbf x,t)\in  \mathbb{R}$ can be written as: 
\begin{equation}
\begin{aligned}
J_1(\mathbf x,t) &= r_1 a p(x_1,x_2,t), \\
J_2(\mathbf x,t) &= r_2 (x_1+1) p(x_1+1,x_2,t), \\
J_3(\mathbf x,t) &= r_3 b p(x_1,x_2,t), \\
J_4(\mathbf x,t) &= r_4 (x_2+1) p(x_1,x_2+1,t),\\
J_5(\mathbf x,t) &= r_5 \frac {(x_1-1) (x_1 -2) x_2}{2}\\
&\times p(x_1-1,x_2+1,t), \\
J_6(\mathbf x,t) &= r_6 \frac {(x_1-1) x_1 (x_1+1) }{6}\\
&\times p(x_1+1,x_2-1,t).
\label{eqn:f_sch}
\end{aligned}
\end{equation}
We have the single-reactional fluxes $J_{5} (\mathbf x,t)  = 0$ and $J_6 (\mathbf x,t) = 0$ on the boundary with either $\mathbf x=(0,0)$ or $\mathbf x=(1,0)$, where reactions $R_5$ and $R_6$ cannot happen. The other single-reactional fluxes are as given above and do not vanish at the boundaries.

The single-reactional velocity $v_k (\mathbf x,t) \in  \mathbb{R}$ can be written as:
$v_k(\mathbf x,t) = J_k(\mathbf x,t)/p(\mathbf x,t).$

\textit{Discrete Partial Derivative.} 
The imposed ordering of the microstates has
$\mathbf x \prec \mathbf x  + \mathbf s_1$, $\mathbf x \prec \mathbf x  - \mathbf s_2$,  $\mathbf x \prec \mathbf x  + \mathbf s_3$, $\mathbf x \prec \mathbf x  - \mathbf s_4$ , $\mathbf x \prec \mathbf x  + \mathbf s_5$, and $\mathbf x \prec \mathbf x  - \mathbf s_6$.
According to Eqs.~(\ref{eqn:dder1})--~(\ref{eqn:dder2}), the derivatives $ {\Delta J_{k} ( \mathbf x, t)}/ {\Delta  \mathbf x_k} $ of the single-reactional fluxes  $J_k$ can be written as:
\begin{eqnarray*}
\frac {\Delta J_{1} (\mathbf x, t)} {\Delta \mathbf x_1} &&=J_{1} (\mathbf x, t) - J_{1} (\mathbf x - \mathbf s_1, t) =r_1 a p(x_1,x_2,t) - r_1 a p(x_1-1,x_2,t), \nonumber \\
\frac {\Delta J_{2} (\mathbf x , t)} {\Delta \mathbf x_2 } && = -(J_{2} (\mathbf x, t) - J_{2} (\mathbf x + \mathbf s_2, t))= - ( r_2 (x_1+1) p(x_1+1,x_2,t) - r_2 x_1 p(x_1,x_2,t)), \nonumber \\
\frac {\Delta J_{3} (\mathbf x, t)} {\Delta \mathbf x_3} &&= J_{3} (\mathbf x, t) - J_{3} (\mathbf x - \mathbf s_3, t) = r_3 b p(x_1,x_2,t) - r_3 b p(x_1,x_2-1,t)), \nonumber \\
\frac {\Delta J_{4} (\mathbf x, t)} {\Delta \mathbf x_4} &&= -(J_{4} (\mathbf x, t) - J_{4} (\mathbf x + \mathbf s_4, t))= - (r_4 (x_2+1) p(x_1,x_2+1,t) - r_4 x_2 p(x_1,x_2,t)), \nonumber \\
\frac {\Delta J_{5} (\mathbf x, t)} {\Delta \mathbf x_5} &&= J_{5} (\mathbf x, t) - J_{5} (\mathbf x - \mathbf s_5, t) = r_5 \frac {x_1 (x_1 -1) x_2} {2} p(x_1,x_2,t) \nonumber \\
&&-r_5 \frac {(x_1-1) (x_1-2) x_2} {2} px_1-1,x_2+1,t)/2,\nonumber\\
\frac {\Delta J_{6} (\mathbf x, t)} {\Delta \mathbf x_6} &&= -(J_{6} (\mathbf x, t) - J_{6} (\mathbf x + \mathbf s_6, t)) = -(r_6 \frac {(x_1-1) x_1 (x_1+1)} {6} p(x_1+1,x_2-1,t) \nonumber \\
&&- r_6 \frac {x_1 (x_1 -1) (x_1-2)} {6} p(x_1,x_2,t)). \nonumber
\end{eqnarray*}

\textit{Total Reactional Flux and Velocity, Discrete Divergence, and Continuity Equation.} Following  Eq.~(\ref{eqn:je}), the total reactional flux
$\mathbf J_r(\mathbf x,t) \in  \mathbb{R}^{6}$ is:
\begin{eqnarray}
\mathbf J_r(\mathbf x,t)=(J_1(\mathbf x ,t), \,J_2(\mathbf x,t),\, J_3(\mathbf x,t), J_4(\mathbf x,t),\, J_5(\mathbf x,t), \,J_6(\mathbf x,t)),\nonumber
\end{eqnarray}
where $\{J_k(\mathbf x, t)\}$ are as specified in Eq.~(\ref{eqn:f_sch}). The total reactional velocity $\mathbf v_r(\mathbf x, t) \in  \mathbb{R}^{6}$ is:
$
\mathbf v_r(\mathbf x,t) = \mathbf J_r(\mathbf x,t)/ p(\mathbf x,t).
$

The discrete divergence $\nabla_{d} \cdot \mathbf J_r (\mathbf x, t)$ of the r-flux $\mathbf J_r(\mathbf x, t) \in \mathbb{R}^{6}$ 
over the discrete increments $\mathbf s_k$ can be written as:
\begin{equation}
\begin{split}
\nabla_{d} \cdot \mathbf J_r (\mathbf x, t) = \sum\limits_{k=1}^{6} {\frac {\Delta J_{k} (\mathbf x,t)} {\Delta \mathbf x_k}}.
\label{eqn:schn_c}
\end{split}
\end{equation}

The r-flux $\mathbf J_r (\mathbf x, t)$ indeed satisfies the continuity equation, as we have
$\nabla_{d} \cdot \mathbf J_r (\mathbf x, t)=-\partial p(\mathbf x,t)/\partial t$ from Eqs.~(\ref{eqn:coneq}), (\ref{eqn:schn_dcme11}), and (\ref{eqn:schn_c})

\textit{Stoichiometry projection and single-reactional species flux.} The single-reactional flux $J_k (\mathbf x,t)$ along the direction of reaction $R_k$ can be decomposed into components of individual species using the predetermined stoichiometry $\mathbf s_{k} = (s_{k}^{1},\, s_{k}^{2})$. 
The $x_1$ and $x_2$ components of \textit{stoichiometric projections} of  $J_{k} (\mathbf x,t)$ are listed in Table~\ref{table:2}.
\begin{table*}
\caption{Schnakenberg system reactional flux stoichiometry projections}
\begin{tabular} {|c | c | c|} 
\hline
Reaction & $J^{1}_{k} (x_1,x_2,t) = s_{k}^{1} J_k(x_1,x_2,t)$ & $J^{2}_{k} (x_1,x_2,t)=s_{k}^{2} J_k(x_1,x_2,t)$ \\ 
\hline
$R_1$ &  $r_1 a p(x_1,x_2,t)$ & $0$\\ 
$R_2$ &  $-r_2 (x_1+1) p(x_1+1,x_2,t)$ & $0$\\ 
$R_3$ &  $0$ & $r_3 b p(x_1,x_2,t)$\\ 
$R_4$ &  $0$ & $-r_4 (x_2+1) p(x_1,x_2+1,t)$\\ 
$R_5$ &  $r_5 \frac {(x_1-1) (x_1 -2) x_2} {2} p(x_1-1,x_2+1,t)$ & $-r_5 \frac {(x_1-1) (x_1 -2) x_2} {2} p(x_1-1,x_2+1,t)$\\ 
$R_6$ &  $-r_6 \frac {(x_1-1) x_1 (x_1+1) (x_1-2)} {6} p(x_1+1,x_2-1,t)$ & $r_6 \frac {(x_1-1) x_1 (x_1+1) (x_1-2)} {6} p(x_1+1,x_2-1,t)$\\ 
\hline
\end{tabular}
\label{table:2}
\end{table*}
The single-reactional species flux is formed as follows: 
\begin{equation}
\mathbf J_k (\mathbf x,t) \equiv (J^{1}_{k} (\mathbf x,t), \, J^{2}_{k} (\mathbf x,t)),\, \quad k=1,\ldots,6,
\label{eqn:f_sch2}
\end{equation}
where $J^{1}_{k} (\mathbf x,t)$ and $J^{2}_{k} (\mathbf x,t)$ listed in Table~\ref{table:2}.
The single-reactional species velocity $\mathbf v_k (\mathbf x,t) \in \mathbb{R^2}$ is  $\mathbf v_k (\mathbf x,t)  \equiv \mathbf J_k (\mathbf x, t) / p (\mathbf x, t).$

\textit{Total Species Flux and Velocity.} Following Eqs.~(\ref{eqn:fl2})--(\ref{eqn:v}), the total flux $\mathbf J_s (\mathbf x,t) \in \mathbb{R}^2$ is
$\mathbf J_s (\mathbf x,t)  =  \sum\limits_{{\text{k = 1}}}^{\text{m}} \mathbf J_k (\mathbf x, t),$
where $\{ \mathbf J_k \}$ as specified in Eq.~(\ref{eqn:f_sch2}).
The total species velocity $\mathbf v_s (\mathbf x,t) \in \mathbb{R}^2$  is:
$
\mathbf v_s (\mathbf x,t)=\mathbf J_s (\mathbf x,t)/ p(\mathbf x,t).
$

\textit{Overall Behavior of Schnakenberg System.}  We examine the
behavior of the Schnakenberg system with $(a, b) = (10, 50)$ under two
initial conditions, namely, that of the uniform distribution and
$\left. {p(\mathbf x = (0,0))} \right|_{t = 0}=1 $. We computed the
time-evolving probability landscape $p=p(\mathbf x,t)$ using the ACME
method~\cite{cao2016state,cao2016accurate}.

\begin{figure*}[thp]
\centering
\includegraphics[scale=1.00]{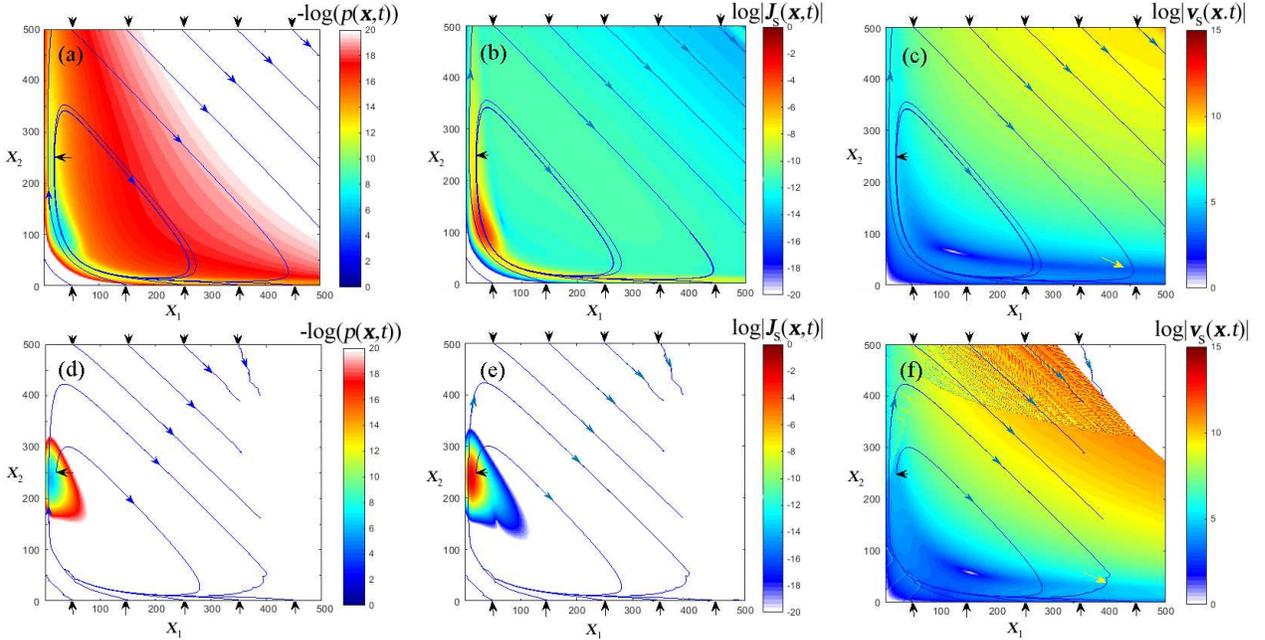}
\caption{ 
The time-evolving probability landscape, flux, and velocity of
probability mass in the Schnakenberg system with $(a,\ b) = (10,\ 50)$
at $t=0.5$,
starting from the uniform distribution (a--c) and from  
the initial conditions of 
 $\left. {p(\mathbf x = (0,0))} \right|_{t = 0}=1$ (d--f). 
a) and d):  the probability landscape in $-\log(p(\mathbf x,t))$;  
b) and e): the corresponding value of flux in $\log|\mathbf J_s(\mathbf x,t)|$; 
c) and f): the $\log$ absolute value of velocity $\log |\mathbf v_s(\mathbf x,t)|$.
}
\label{fig:st}
\end{figure*}
\begin{figure*}[thp]
\centering
\includegraphics[scale=1.00]{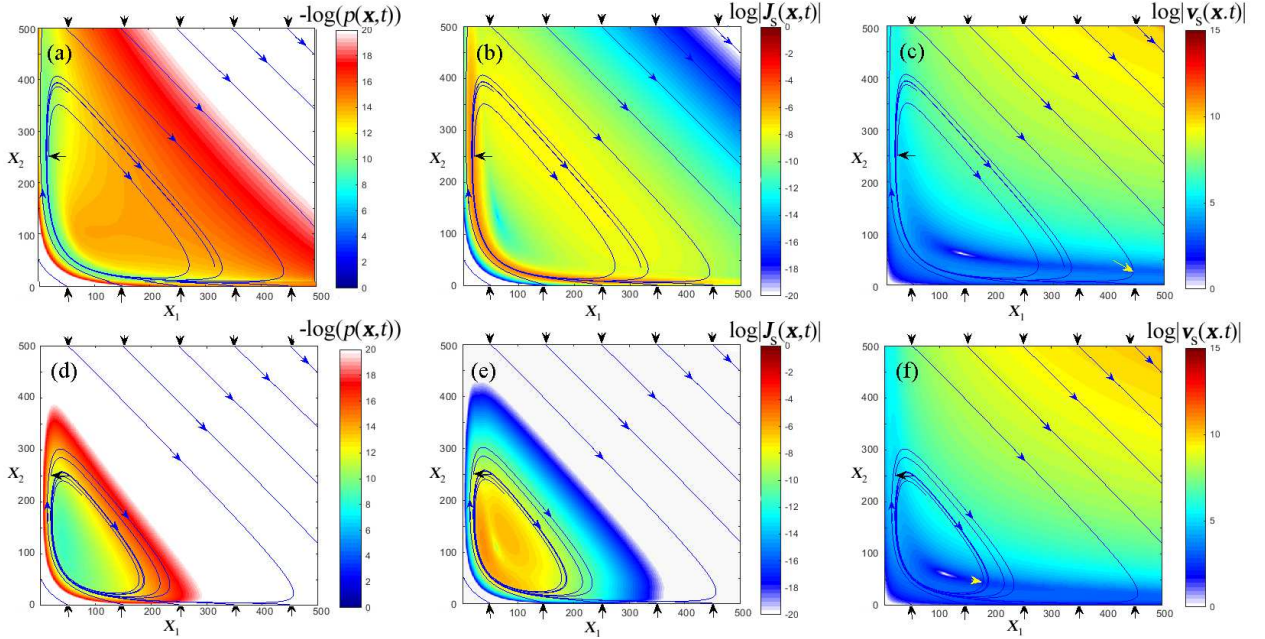}
\caption{
The steady-state probability landscape, flux, and velocity of
probability mass in the Schnakenberg system with $(a,\ b) = (10,\ 50)$
(a--c) and $(a, b) = (20, 40)$ (d--f). 
a) and d):  the probability landscape in $-\log(p(\mathbf x,t))$ ;
b) and e): the corresponding values of flux in $\log|\mathbf J_s(\mathbf x,t)|$; 
c) and f): the $\log$ absolute value of velocity $\log |\mathbf v_s(\mathbf x,t)|$.}
\label{fig:s_ss}
\end{figure*}

For the uniform distribution, the probability
landscape in $-\log p(\mathbf x, t)$ at time $t=0.5$ is shown in Figure~\ref{fig:st}a, where high
probability regions are in blue.  Its overall shape takes the
form of closed valley, which is similar to an earlier study based on a Fokker-Planck
model~\cite{xu2012energy}.
The trajectories of the flux field $\mathbf{J}_s
(\mathbf x, t)$ at time $t=0.5$ in the space of the copy-numbers from different starting locations (marked by black arrows at top and bottom) 
are shown in blue on Fig.~\ref{fig:st}-~\ref{fig:s_ss}. These trajectories depict the directions of the movement of the probability mass at different locations after traveling from the starting points.
The heatmaps of the flux in $\log
|J_s(\mathbf x,t)|$ and the velocity in $\log |v_s(\mathbf x,t)|$ are
shown in Fig.\ \ref{fig:st}b and Fig.~\ref{fig:st}c, respectively.
The flux lines are closed curves and are overall smooth.  These closed
flux lines reflect the oscillatory nature of the reaction system.
The velocity has larger values at locations where the flux
trajectories are straight lines (green and yellow region in the upper
right corner, Figure~\ref{fig:st}c), but drops significantly when the
trajectories make down-right turns (light and dark blue in the lower
right corner, marked with an yellow arrow).

For the initial conditions of $\left. {p(\mathbf x = (0,0))}
\right|_{t = 0}=1$, $-\log p(\mathbf x, t)$ at time $t=0.5$ is shown in 
Figure~\ref{fig:st}d,
where high
probability regions (blue)  is located at a small neighborhood around $\mathbf x=(0,\,250)$.
The heatmaps of the flux in $\log
|J_s(\mathbf x,t)|$ and the velocity in $\log |v_s(\mathbf x,t)|$ are
shown in Fig.\ \ref{fig:st}e and Fig.~\ref{fig:st}f, respectively.
The flux lines are closed curves and are overall smooth.  
The oscillating flux lines appear again (Figs~\ref{fig:st}d--~\ref{fig:st}f), but not all form closed
curves.  Specifically, all flux lines which
start at the upper region ($x_2=500$) become broken-off in the mid-region,
where the probability mass becomes negligible, resulting in negligible
flux as well, with its absolute value close to be zero.
The maximum of the flux is reached at the
peak of the probability landscape (Figure~\ref{fig:st}e). 
The heatmap of the probability velocity exhibits a similar pattern as
that of uniform distribution (Figure~\ref{fig:st}f
vs.\ Figure~\ref{fig:st}c).  The color palettes encoding the values
of the velocity $\log |v_s(\mathbf x,t)|$ are not-smooth
(Figure~\ref{fig:st}f). This is likely due to small numerical values of
probability in this region.

We then examined the steady state behavior of the system at two
conditions of the copy numbers of species $A$ and $B$: $(a,b) =
(10,50)$ and $(a, b) = (20, 40)$.  
The probability landscape in $-\log(p(\mathbf x,t))$ for $(a,b) =
(10,50)$ shown in Fig.~\ref{fig:s_ss}a exhibits similar shape to that of
Fig~\ref{fig:st}.  The probability values are higher in locations near
the left ($x_1 = 0$) and lower ($x_2 = 0$) boundaries.
The flux lines (Fig.~\ref{fig:s_ss}a-~\ref{fig:s_ss}c) move from the upper left
corner to the lower right corner, and then make sharp right turns until
reaching the neighborhood near the origin.  Subsequently, they make right turns again and move
upward, until the cycles are closed.  These closed flux curves move along
the contours on the probability landscape.
The absolute values of the flux (Fig.~\ref{fig:s_ss}b) are largest near
the boundaries of the probability surfaces ($x_1 = 0$ and $x_2=0$, red/orange colored ridge) and
nextly along the flux lines on the diagonal.  The flux has small values in the
region above the diagonal (cyan and blue).  The heatmap of the
velocity (Fig.~\ref{fig:s_ss}c) exhibit a different pattern, with its
value dropping significantly in the small blue arch (see region pointed
by the yellow arrow), where flux lines make turns in the lower region.

The probability landscape in $-\log(p(\mathbf x,t))$ for $(a,b) =
(20,40)$ is shown in Fig.~\ref{fig:s_ss}d. 
While exhibiting overall
similar 
pattern to that of $(a,b) = (10,50)$,
the high probability
regions is more concentrated in locations near the lower-left
 (Fig.~\ref{fig:s_ss}d).
The flux lines (Fig.~\ref{fig:s_ss}d--f) 
are similar to those of $(a,b) = (10,50)$ corner, but
oscillate around much smaller contour, where $x_1 \le
200$ and $x_2 \le 300$.  
The close cycles of flux
lines also move along the contours on the probability landscape.

The results obtained here are generally consistent with that obtained
using a Fokker-Planck flux model computed from a landscape constructed using Gillespie simulations~\cite{gillespie1977exact, xu2012energy}. For example, the
directions of the flux lines are the same.  However, there are some
differences.
While the flux lines from the Fokker-Planck model exhibit oscillating
behavior even in the boundary regions where $x_1 <2$ or $x_2 <2$,
where reactions $R_5$ and $R_6$ cannot occur hence no oscillating flux
are physically possible.  No such inconsistency exist in our model.
Furthermore, the system considered here is much larger, with hundreds
of copies of $X_1$ and $X_2$ involed, whereas $<10$ copies of $X_1$
and $X_2$ were considered in~\cite{xu2012energy}.

\section{Conclusion}
\label{Conclusion}
In this study, we introduce
new formulations of discrete flux and discrete velocity for an arbitrary mesoscopic reaction system. Specifically, we redefine the derivative and divergence operators based on the 
discrete nature of chemical reactions. We then
introduce the
discrete form of continuity equation for the systems of reactions. We define two types of discrete flux, with their relationship specified. 
The reactional discrete flux satisfies the continuity equation and
 describes the behavior of the system evolving along directions of 
 reactions. The species flux directly describes the dynamic behavior of the
reactions such as the transfer of probability mass in the state space. Our discrete flux model enables the construction of the global time-evolving and steady-state flow-maps of fluxes 
   in all directions at every microstate. Furthermore, it can be used to tag the fluxes of outflow and inflow of probability mass as reactions proceeds.
In addition,
 we can now impose boundary conditions,
 allowing exact quantification of 
 vector fields of the discrete flux and discrete velocity anywhere in the discrete state space, without the difficulty of enforcing artificial reflecting conditions at the
boundaries~\cite{ceccato2018remarks}.
  We note that the accurate construction of the discrete probability flux, velocity,
and their global flow-maps requires the accurate calculation of the time-evolving probability landscape of
the reaction network.
This is made possible by using the
recently developed ACME method~\cite{cao2016accurate,cao2016state}.
  
As a demonstration, we computed the time-evolving probability flux and velocity fields for
 three model systems, namely, the birth-death process, the bistable
 Schl\"ogl model, and the oscillating Schnakenberg system.  
We showed
 how flux and velocities converge to zero when the system reaches
 the steady-state in the birth-death process and the Schl\"ogl models.  
We also showed that the flux and velocity trajectories in the
Schnakenberg system converge to the oscillating contours of the steady-state
probability landscape, similar to an earlier
study~\cite{xu2012energy}, although there are important differences.
Overall, the general framework of discrete flux and velocity and the
methods introduced here can be applied to other networks and
dynamical processes involving stochastic reactions.
These applications can be useful in quantification of dynamic changes of
probability mass, identification as well as characterization of
mechanism where movement of probability mass drives the
system towards the steady-state.
They may also aid in our understanding of the mechanisms that determined 
 the non-equilibrium steady state of many reaction systems.

\section{Acknowledgments}
\label{Acknowledgments} Support from NIH R35 GM127084 and NSF  DMS-1714401 is gratefully
acknowledged.

\bibliographystyle{unsrt}

\end{document}